\documentstyle[amssymb,prl,aps,epsfig,twocolumn]{revtex}
\begin{document}
\draft

\twocolumn[\hsize\textwidth\columnwidth\hsize\csname@twocolumnfalse\endcsname

\title{Holonomic Quantum Computation Using Rf-SQUIDs Coupled Through A
Microwave Cavity}
\author{P. Zhang $^{1,2}$, Z. D. Wang $^{1}$, J. D. Sun $^{1}$ and C. P. Sun
$^{2}$}
\address{$^{1}$Department of Physics, the University of Hong Kong, Hong Kong, China\\
$^{2}$Institute of Theoretical Physics, the Chinese Academy of
Science, Beijing, 100080, China}
\maketitle

\begin{abstract}
We propose a new scheme to realize holonomic quantum computation with
rf-SQUID qubits in a microwave cavity. In this scheme associated with the
non-Abelian holonomies, the single-qubit gates and a two-qubit control-Phase
gate as well as a control-NOT gate can be easily constructed by tuning
adiabatically the Rabi frequencies of classical microwave pulses coupled to
the SQUIDs. The fidelity of these gates is estimated to be possibly higher
than 90 \% with the current technology.
\end{abstract}

\pacs{PACS number: 03.67.Lx, 03.65.Vf, 85.25.Cp }]

%\section{Introduction}

%Since the discovery of Berry's phase \cite{berry}, the geometric
%effects in quantum evolutions have been widely studied
%\cite{phase}. In 1999, the all-geometric approaches of quantum
%computation \cite{qc} are
Since the proposal of holonomic quantum computation \cite{holonomic},
research on quantum gates based on
%single-qubit and two-qubit gates are realized via
Abelian or Non-Abelian geometric phase shifts has attracted significant
interests both experimentally and theoretically\cite%
{ekert,jj2,duan,gate1,SZhu1,nmr2,jj,sun,im2}. It is believed that these
quantum gates could be inherently robust against some local perturbations
since the Abelian or Non-Abelian geometric phases(holonomies) depend only on
the geometry of the path executed. On the other hand, quantum information
processing using Josephson-junction systems coupled through a microwave
cavity has been paid particular attentions recently\cite%
{mail1,mail2,prb,zdwang2,han2,zdwang,gao,you,liu}. Nevertheless,
how to realize the holonomic quantum computation using
superconducting quantum interference devices(SQUIDs) in a cavity
has not been addressed.

In this paper, we propose a novel scheme to achieve holonomic quantum
computation using SQUIDs in a cavity. Based on the non-Abelian holonomies,
two non-commutating single-qubit gates and a two-qubit control-Phase gate as
well as a control-NOT gate are realized by tuning adiabatically the Rabi
frequencies of classical microwave pulses coupled to the SQUIDs. The
distinct advantages of the present scheme may be summarized as follows.
%(i)The position of the SQUID qubits can be fixed in the cavity.
(i) The energy spectrum of each SQUID qubit may be adjusted by changing the
bias field; (ii) the strong coupling limit $g^{2}>>\left( \gamma \kappa
\right) $ may be easily realized, where $g$ is the coupling coefficients
between the SQUID qubit and the cavity field, $\kappa $ the life time of the
photon in the cavity and $\gamma $ the life time of the excited state of the
SQUID qubit; (iii) the decoherence caused by the external environment can be
significantly suppressed; (iv) the fidelity of these gates may be higher
than 90 \% with the current technology.

% \section{Single-Qubit Operation}

We consider an rf-SQUID (with junction capacitance $C$ and loop inductance $%
L $) in an microwave cavity.  The Hamiltonian of the rf-SQUID can
be written
as \cite{han2,han3}%
\begin{equation}
H_{s}=\frac{Q^{2}}{2C}+\frac{\left( \Phi -\Phi _{x}\right) ^{2}}{2L}%
-E_{J}\cos \left( 2\pi \frac{\Phi }{\Phi _{0}}\right)  \label{a}
\end{equation}%
where $E_{J}$ is the maximum Josephson coupling energy, $\Phi _{x}$ the
external magnetic flux and $\Phi _{0}=h/2e$ the flux quantum. The conjugate
variables of this system are the total charge $Q$ and the magnetic flux $%
\Phi $ which satisfy $\left[ \Phi ,Q\right] =i\hbar $. It is well
known that the Hamiltonian of Eq. (\ref{a}) is quite similar to
that of a particle moving in a double well potential. By changing
the device parameters $C$, $L$ and the control parameters $E_{J}$,
$\Phi _{x}$, one can control the structure of energy levels in the
SQUID.

Let us address a (3+1)-type system with three lowest levels ($\left\vert
a_{0}\right\rangle $, $\left\vert a_{1}\right\rangle $, $\left\vert
g\right\rangle $) and an excited level ($\left\vert e\right\rangle $) in the
SQUID (see Fig. 1). In the system, the $\left\vert g\right\rangle
\leftrightarrow \left\vert e\right\rangle $ transition with energy level
difference $\omega _{eg}$ is coupled to an one-mode cavity field with
frequency $\omega _{c}$ and the $\left\vert a_{i}\right\rangle
\leftrightarrow \left\vert e\right\rangle $ transition with the energy level
difference $\omega _{ei}$ ($i=0,1$) is coupled to the classical microwave
pulse with the magnetic component as ${\bf B}_{i}\left( {\bf r},t\right)
\cos \left( \omega _{0}t\right) $ where $\omega _{i}$ is the energy
difference between the states $\left\vert a_{i}\right\rangle $ and $%
\left\vert e\right\rangle $ and ${\bf B}_{i}$ can be adiabatically changed.
We may ensure that the "{\it 3-photon resonance}" condition, i.e. $\omega
_{eg}-\omega _{c}=\omega _{ei}-\omega _{i}=\Delta $, is satisfied. In the
interaction picture, the Hamiltonian of the system can be written as%
\begin{eqnarray}
H &=&\Delta \left\vert e\right\rangle \left\langle e\right\vert +\Omega
_{0}\left( t\right) \left\vert e\right\rangle \left\langle a_{0}\right\vert
+\Omega _{1}\left( t\right) \left\vert e\right\rangle \left\langle
a_{1}\right\vert  \label{b} \\
&&+ga\left\vert e\right\rangle \left\langle g\right\vert +h.c.  \nonumber
\end{eqnarray}%
where $a$ ($a^{\dagger }$) is the photon annihilation (creation) operator of
the cavity field. Here the coupling coefficients $g$ can be written as $%
g=L^{-1}\sqrt{\omega _{c}/\left( 2\mu _{0}\hbar \right) }\left\langle
g\right\vert \Phi \left\vert e\right\rangle \int_{S}{\bf B}_{c}\left( {\bf r}%
\right) \cdot d{\bf S}$ where $\mu _{0}$ is the vacuum permeability, $S$ the
surface bounded by the SQUID ring and ${\bf B}_{c}\left( {\bf r}\right) $
the magnetic component of the cavity field\cite{han2}. The Rabi frequencies $%
\Omega _{i}\left( t\right) $ ($i=0,1$) are proportional to the matrix
elements $\left\langle a_{i}\right\vert \Phi \left\vert e\right\rangle $
\cite{han2}. It is pointed out that the detuning $\Delta $ of the SQUID can
be adjusted by changing the bias field.

\begin{figure}[h]
\begin{center}
\includegraphics[width=4cm,height=4cm]{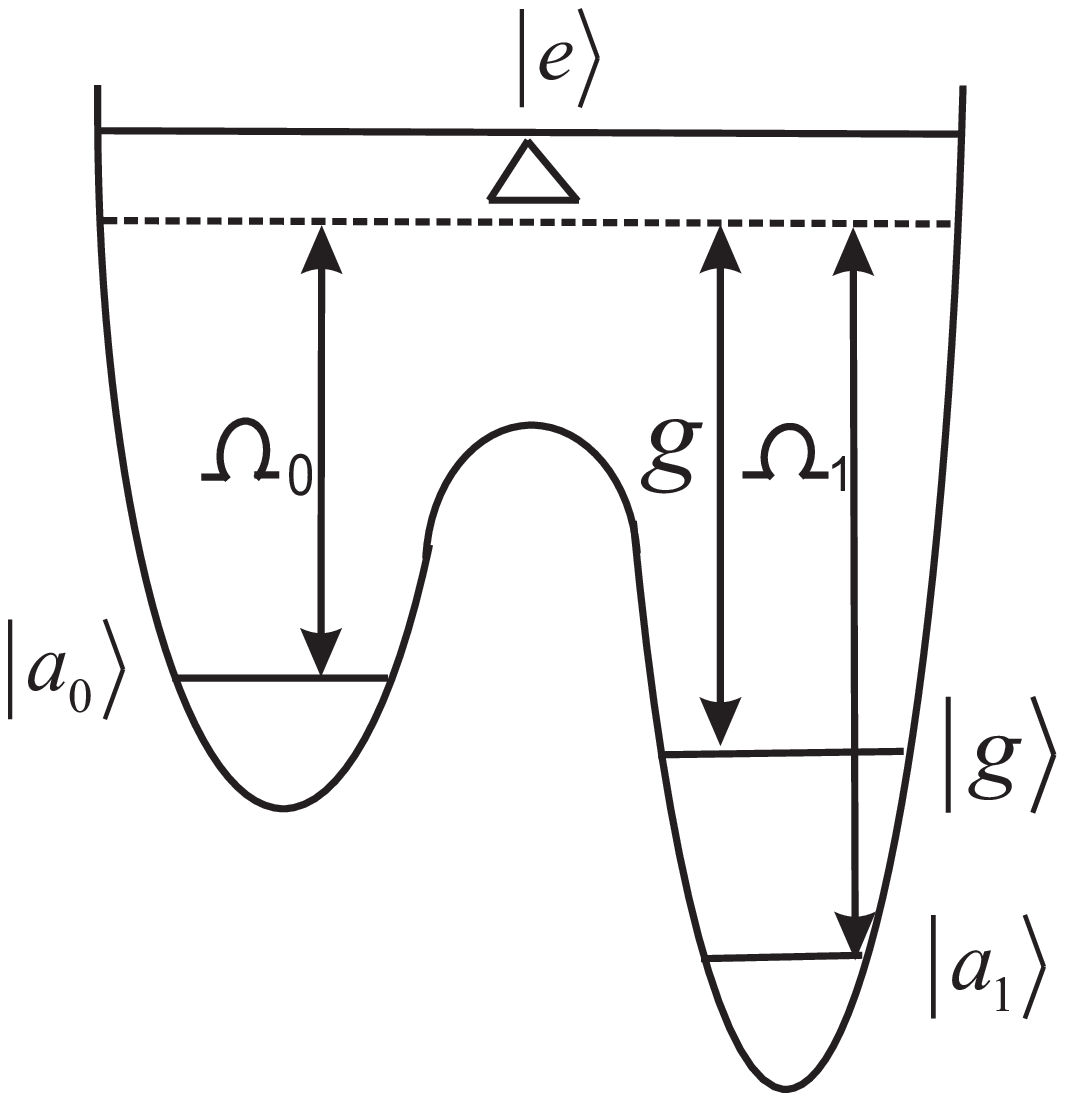} \vspace{0.3cm}
\end{center}
\caption{A schematic diagram of the energy level in the SQUID coupled to the
single mode cavity field (with coupling constant $g$) and two microwave
pulses (with coupling constants $\Omega _{0}$ and $\Omega _{1}$). The
3-photon resonance condition is satisfied and $\Delta $ is the detuning.}
\end{figure}

Regarding single-qubit operations, our scheme is similar to that proposed in
\cite{duan,gate1}. We choose the states $\left\vert a_{0}\right\rangle $ and
$\left\vert a_{1}\right\rangle $ as the computational basis and $\left\vert
g\right\rangle $ as an ancillary state. When $\Omega _{0}=\Omega _{1}=0$,
the states $\left\vert a_{i}\right\rangle \left\vert 0\right\rangle _{c}$ ($%
i=0,1$) span an eigenspace of the Hamiltonian (\ref{b}) with zero
eigenvalue., where, $\left\vert 0\right\rangle _{c}$ is the vacuum state of
the cavity field. If the cavity is cooled to zero temperature and the
quantum gate operations are switched off, i.e., the Rabi frequencies of all
the classical microwave pulses are set to zero, the state $\alpha \left\vert
a_{0}\right\rangle +\beta \left\vert a_{1}\right\rangle $ of the qubit is
isolated from the state of the cavity and does not change with time. When
the Rabi frequencies $\Omega _{0}$ and $\Omega _{1}$ change adiabatically
along a close path $C$ in the parameter space $M$ and return the point
corresponding to $\Omega _{0}=\Omega _{1}=0$ (we refer to this point as $O$%
), an initial state $\left\vert \Psi _{0}\right\rangle =\left( \alpha
\left\vert a_{0}\right\rangle +\beta \left\vert a_{1}\right\rangle \right)
\left\vert 0\right\rangle _{c}$ of the qubit-cavity composite system evolves
according to the rule $\left\vert \Psi _{0}\right\rangle \rightarrow U\left(
C\right) \left\vert \Psi _{0}\right\rangle $ \cite{gate1,zee}. Here, $%
U\left( C\right) =P\exp \int_{C}A$ is the non-Abelian holonomy associated
with the path $C$ and $A=\sum_{\mu }A_{\mu }d\lambda _{\mu }$ is the $%
U\left( 2\right) $-valued connection expressed as $A_{\mu
}^{ij}=\left\langle D_{i}\left( \lambda \right) \right\vert \frac{\partial }{%
\partial \lambda _{\mu }}\left\vert D_{j}\left( \lambda \right)
\right\rangle $ where $\left\{ \lambda _{\mu }\right\} $ are the coordinates
of the parameter space $M$ and $\left\vert D_{i}\left( \lambda \right)
\right\rangle $ ($i=0,1$) the basis of the eigenspace of the Hamiltonian (%
\ref{b}) with zero eignevalue (hereafter we refer to the basis as dark
states). Since the holonomy $U\left( C\right) $ is a unitary transformation
in the space spanned by the states $\left\vert a_{0}\right\rangle \left\vert
0\right\rangle _{c}$ and $\left\vert a_{1}\right\rangle \left\vert
0\right\rangle _{c}$, it can actually be considered as a unitary
transformation that only acts on the qubit state which is the superposition
of $\left\vert a_{0}\right\rangle $ and $\left\vert a_{1}\right\rangle $.

Without loss of generality, we assume that the coupling coefficient $g$ is
real and positive, and choose $\Omega _{0}=g\tan \left( \xi \right) e^{i\phi
_{0}}$and $\Omega _{1}=g\tan \left( \theta \right) \sec \left( \xi \right)
e^{i\phi _{1}}$ where $\theta ,\xi \in \left[ 0,\frac{\pi }{2}\right) $ and $%
\phi _{i}\in \left[ 0,2\pi \right) $ $\left( i=0,1\right) $. We take the
angles $\xi $, $\theta $, $\phi _{0}$ and $\phi _{1}$ as the coordinates of
the parameter space $M$. The dark states of this invariant subspace spanned
by the states $\{\left\vert a_{i}\right\rangle \left\vert 0\right\rangle
_{c} $, $\left\vert g\right\rangle \left\vert 1\right\rangle _{c}$, $%
\left\vert e\right\rangle \left\vert 0\right\rangle _{c}\}$ of the
Hamiltonian (\ref{b}) can be written as the vector functions in $M$: $%
\left\vert D_{0}\right\rangle =\cos \left( \xi \right) \left\vert
a_{0}\right\rangle \left\vert 0\right\rangle _{c}-\sin \left( \xi \right)
e^{i\phi _{0}}\left\vert g\right\rangle \left\vert 1\right\rangle _{c}$ and $%
\left\vert D_{1}\right\rangle =-\sin \left( \theta \right) \sin \left( \xi
\right) e^{i\left( \phi _{1}-\phi _{0}\right) }\left\vert a_{0}\right\rangle
\left\vert 0\right\rangle _{c}-\sin \left( \theta \right) \cos \left( \xi
\right) e^{i\phi _{1}}\left\vert g\right\rangle \left\vert 1\right\rangle
_{c}+\cos \left( \theta \right) \left\vert a_{1}\right\rangle \left\vert
0\right\rangle _{c}.$ We wish to point out that our choices of the
coordinates and dark states are quite different from those in Refs. \cite%
{duan,gate1}. Here, the dark states $\left\vert D_{0}\right\rangle $ and $%
\left\vert D_{1}\right\rangle $ are single-valued at the point $O$ ($\theta
=\xi =0$), {\it e.g.} $\left\vert D_{i}\left( O\right) \right\rangle
=\left\vert a_{i}\right\rangle \left\vert 0\right\rangle _{c}$.

It is well known that any single-qubit gate operation can be decomposed into
the product of rotations about axies $z$ and $y$: $R_{z}\left( \varphi
\right) =e^{i\varphi \sigma _{z}}$ and $R_{y}\left( \varphi \right)
=e^{i\varphi \sigma _{y}}$, where, $\varphi$ is the angle, $\sigma _{z}$,
and $\sigma _{y}$ are Pauli matrices defined as $\sigma _{y}=i\left(
\left\vert a_{1}\right\rangle \left\langle a_{0}\right\vert -\left\vert
a_{0}\right\rangle \left\langle a_{1}\right\vert \right) $ and $\sigma
_{z}=\left( \left\vert a_{0}\right\rangle \left\langle a_{0}\right\vert
-\left\vert a_{1}\right\rangle \left\langle a_{1}\right\vert \right) $.
Therefore, we need only to show the realization of $R_{z}\left( \varphi
\right) $ and $R_{y}\left( \varphi \right) $. To realize the gate $%
R_{y}\left( \varphi \right) $, we let the phases $\phi _{0}=\phi _{1}=0$.
The $U\left( 2\right) $ valued connections can be derived as $A_{\xi
}=-i\sin \left( \theta \right) \sigma _{y}$, and $A_{\theta }=0$. After an
adiabatic evolution along a closed path $C$ in the parameter space $M$, the
related unitary transformation (holonomy) is just the rotation about $y$
axis $U\left( C\right) =\exp \left[ i\varphi \left( C\right) \sigma _{y}%
\right] $, where the angle $\varphi \left( C\right) =-\left( \int_{C}\sin
\left( \theta \right) d\xi \right) $ is dependent of the loop $C$. To
achieve the gate $R_{z}\left( \varphi \right) $, we can set $\phi _{0}=\xi
=0 $ (i.e. $\Omega _{0}=0$) and change $\theta $ and $\phi _{1}$
adiabatically. In this case, the non-zero connection is $A_{\phi _{1}}=i%
\frac{1}{2}\sin ^{2}\left( \theta \right) \left( 1-\sigma _{z}\right) $. As
a result, the holonomy associated with a close path $C$ can be written as $%
U\left( C\right) =e^{-i\chi \left( C\right) }e^{i\chi \left( C\right) \sigma
_{z}}$ where $\chi \left( C\right) =-\frac{1}{2}\int \sin ^{2}\left( \theta
\right) d\phi _{1}$. It is easy to see that $U\left( C\right) $ is just a
rotation about $z $ axis up to a global phase.

We now illustrate how to realize the controlled-PHASE gate as well as the
controlled-NOT gate via the non-Abelian holonomy in the present system of
SQUID qubits, which is the main and novel result of the present paper. At
this stage, we consider the (3+1)-type energy-level structure of two
rf-SQUIDs in a microwave cavity. We assume that the $\left\vert
e\right\rangle \longleftrightarrow \left\vert g\right\rangle $ transition of
each SQUID is coupled to the single mode cavity field. Also, we set the
level spacing $\omega _{ei}$ between the states ($\left\vert e\right\rangle $%
, $\left\vert a_{i}\right\rangle ,$ $i=0,1$) to be different. The
transitions $\left\vert e\right\rangle \longleftrightarrow \left\vert
a_{0}\right\rangle $ (or $\left\vert a_{1}\right\rangle $) in both SQUIDs
are coupled to two distinguishable classical microwave pulses with different
frequencies \cite{han2} and the "{\it 3-photon resonance}" condition of each
SQUID is set to be satisfied. The Hamiltonian of such SQUIDs in a microwave
cavity may be written as%
\begin{eqnarray}
H &=&\sum_{l=1,2}\left[ \Omega _{0}^{\left( l\right) }\left\vert
e\right\rangle _{l}\left\langle a_{0}\right\vert +\Omega _{1}^{\left(
l\right) }\left\vert e\right\rangle _{l}\left\langle a_{1}\right\vert +h.c%
\right]  \label{j} \\
&&+\sum_{l=1,2}\left[ g^{\left( l\right) }a\left\vert e\right\rangle
_{l}\left\langle g\right\vert +h.c\right] +\sum_{l=1,2}\Delta ^{\left(
l\right) }\left\vert e\right\rangle _{l}\left\langle e\right\vert .
\nonumber
\end{eqnarray}%
Here, $\Delta ^{\left( l\right) }=\omega _{eg}^{\left( l\right) }-\omega
_{c}^{\left( l\right) }=\omega _{ei}^{\left( l\right) }-\omega _{i}^{\left(
l\right) }$ is the detunning of the $l$-th SQUID and $\Omega _{i}^{\left(
l\right) }$ is the Rabi frequency of the microwave pulse coupled to the
transition $\left\vert e\right\rangle _{l}\leftrightarrow \left\vert
a_{i}\right\rangle _{l}$ of the $l$-th SQUID.

We still choose the states $\left\vert a_{i}\right\rangle _{l}$ $i=0,1$ as
the computational basis of the $l$-th qubit. The two SQUIDs are coupled
indirectly via the single mode cavity field. As in the previous discussions
for the single-qubit gate case, it is seen that when all of the Rabi
frequencies $\Omega _{i}^{\left( l\right) }$ of the classical microwave
pluses are set to zero and the cavity is cooled to zero temperature, the
two-qubit state, which can be written in terms of $\left\vert
a_{i}\right\rangle _{1}\left\vert a_{j}\right\rangle _{2}$, is isolated from
the state of the cavity and does not change with time. In the following, we
show that the holonomy, associated with the adiabatic evolution of $\Omega
_{i}^{\left( l\right) }$ along a closed path $C$ starting from the point $%
\Omega _{i}^{\left( l\right) }=0$ (we also refer to this point as $O$) in
the parameter space $M$, can be used to achieve a controlled-PHASE gate or a
controlled-NOT gate of the two qubits.

We first choose $\Omega _{0}^{\left( l\right) }=g^{\left( 1\right) }\tan %
\left[ \xi ^{\left( l\right) }\right] e^{i\phi _{0}^{\left( l\right) }}$ and
$\Omega _{1}^{\left( l\right) }=g^{\left( 2\right) }\tan \left[ \theta
^{\left( l\right) }\right] \sec \left[ \xi ^{\left( l\right) }\right]
e^{i\phi _{1}^{\left( l\right) }}$ ( $l=1,2$), and take $\xi ^{\left(
l\right) }$, $\phi _{0}^{\left( l\right) }$, $\theta ^{\left( l\right) }$
and $\phi _{1}^{\left( l\right) }$ as the independent coordinates in the
parameter space $M$. To realize the controlled-PHASE gate, we set $\theta
^{\left( l\right) }=\phi _{1}^{\left( l\right) }=\phi _{0}^{\left( 1\right)
}=0$. This means that we set the Rabi frequencies $\Omega _{1}^{\left(
1\right) }$ and $\Omega _{1}^{\left( 2\right) }$ to zero and only change $%
\Omega _{0}^{\left( 1\right) }$ and $\Omega _{0}^{\left( 2\right) }$
adiabatically. It is found that the subspace spanned by the states $%
\left\vert a_{i}\right\rangle _{1}\left\vert a_{j}\right\rangle
_{2}\left\vert 0\right\rangle _{c}$, $\left\vert a_{i}\right\rangle
_{1}\left\vert g\right\rangle _{2}\left\vert 1\right\rangle _{c}$, $%
\left\vert g\right\rangle _{1}\left\vert g\right\rangle _{2}\left\vert
2\right\rangle _{c}$, $\left\vert g\right\rangle _{1}\left\vert
a_{i}\right\rangle _{2}\left\vert 1\right\rangle _{c}$, $\left\vert
e\right\rangle _{1}\left\vert a_{i}\right\rangle _{2}\left\vert
0\right\rangle _{c}$, $\left\vert a_{i}\right\rangle _{1}\left\vert
e\right\rangle _{2}\left\vert 0\right\rangle _{c}$, $\left\vert
e\right\rangle _{1}\left\vert g\right\rangle _{2}\left\vert 1\right\rangle
_{c}$, $\left\vert g\right\rangle _{1}\left\vert e\right\rangle
_{2}\left\vert 1\right\rangle _{c}$ and $\left\vert e\right\rangle
_{1}\left\vert e\right\rangle _{2}\left\vert 0\right\rangle _{c}$ $\left(
i,j=0,1\right) $ is an invariance subspace (we call it as the subspace $I$)
of the Hamiltonian (\ref{j}). After some tedious derivations, the dark
states, i.e. the basis of the eigenspace of Hamiltonian (\ref{j}) with zero
eigenvalue, can be obtained as:
\begin{eqnarray}
\left\vert D_{00}\right\rangle &=&\Lambda _{00}^{-1}[\sin \left[ \xi
^{\left( 1\right) }\right] \sin \left[ \xi ^{\left( 2\right) }\right]
e^{i\phi _{0}^{\left( 2\right) }}\left\vert g\right\rangle _{1}\left\vert
g\right\rangle _{2}\left\vert 2\right\rangle _{c}  \nonumber \\
&&-\sqrt{2}\cos \left[ \xi ^{\left( 1\right) }\right] \sin \left[ \xi
^{\left( 2\right) }\right] e^{i\phi _{0}^{\left( 2\right) }}\left\vert
a_{0}\right\rangle _{1}\left\vert g\right\rangle _{2}\left\vert
1\right\rangle _{c}  \nonumber \\
&&+\sqrt{2}\cos \left[ \xi ^{\left( 1\right) }\right] \cos \left[ \xi
^{\left( 2\right) }\right] \left\vert a_{0}\right\rangle _{1}\left\vert
a_{0}\right\rangle _{2}\left\vert 0\right\rangle _{c}  \nonumber \\
&&-\sqrt{2}\sin \left[ \xi ^{\left( 1\right) }\right] \cos \left[ \xi
^{\left( 2\right) }\right] \left\vert g\right\rangle _{1}\left\vert
a_{0}\right\rangle _{2}\left\vert 1\right\rangle _{c}]  \nonumber \\
\left\vert D_{01}\right\rangle &=&\cos \left[ \xi ^{\left( 1\right) }\right]
\left\vert a_{0}\right\rangle _{1}\left\vert a_{1}\right\rangle
_{2}\left\vert 0\right\rangle _{c}  \label{l} \\
&&-\sin \left[ \xi ^{\left( 1\right) }\right] \left\vert g\right\rangle
_{1}\left\vert a_{1}\right\rangle _{2}\left\vert 1\right\rangle _{c}
\nonumber \\
\left\vert D_{10}\right\rangle &=&\cos \left[ \xi ^{\left( 2\right) }\right]
\left\vert a_{1}\right\rangle _{1}\left\vert a_{0}\right\rangle
_{2}\left\vert 0\right\rangle _{c}  \nonumber \\
&&-\sin \left[ \xi ^{\left( 2\right) }\right] e^{i\phi _{0}^{\left( 2\right)
}}\left\vert a_{1}\right\rangle _{1}\left\vert g\right\rangle _{2}\left\vert
1\right\rangle _{c}  \nonumber \\
\left\vert D_{11}\right\rangle &=&\left\vert a_{1}\right\rangle
_{1}\left\vert a_{1}\right\rangle _{2}\left\vert 0\right\rangle _{c}
\nonumber
\end{eqnarray}%
where $\Lambda _{00}=\sqrt{2-\sin ^{2}\left[ \xi ^{\left( 1\right) }\right]
\sin ^{2}\left[ \xi ^{\left( 2\right) }\right] }$. It is obvious that at the
point $O$ where $\xi ^{\left( 1\right) }=\xi ^{\left( 2\right) }=0$, the
above dark states are single valued: $\left\vert D_{ij}\left( O\right)
\right\rangle =\left\vert a_{i}\right\rangle _{1}\left\vert
a_{j}\right\rangle _{2}\left\vert 0\right\rangle _{c}$ for $i,j=0,1$. The
non-zero elements of $U\left( 4\right) $-valued connections are just $%
A_{\phi _{0}^{\left( 2\right) }}^{00,00}=i\Lambda _{00}^{-2}\left\{ 2-\sin
^{2}\left[ \xi ^{\left( 1\right) }\right] \right\} \sin ^{2}\left[ \xi
^{\left( 2\right) }\right] $ and $A_{\phi _{0}^{\left( 2\right)
}}^{10,10}=i\sin ^{2}\left[ \xi ^{\left( 2\right) }\right] .$ When the
system evolves adiabatically along a closed path $C$ in the parameter space $%
M$ and returns to the point $O$, the associated holonomy can be written as $%
U\left( C\right) =e^{i\eta \left( C\right) \left\vert a_{0}\right\rangle
_{2}\left\langle a_{0}\right\vert }e^{i\phi \left( C\right) \left\vert
a_{0}a_{0}\right\rangle \left\langle a_{0}a_{0}\right\vert }$. Here, the
angles $\eta \left( C\right) $ and $\phi \left( C\right) $ are defined as $%
\phi \left( C\right) =\int \Lambda _{00}^{-2}\left[ 2-\sin ^{2}\left[ \xi
^{\left( 1\right) }\right] -\Lambda _{00}^{2}\right] \sin ^{2}\left[ \xi
^{\left( 2\right) }\right] d\phi _{0}^{\left( 2\right) }$ and $\eta \left(
C\right) =\int \sin ^{2}\left[ \xi ^{\left( 2\right) }\right] d\phi
_{0}^{\left( 2\right) }.$ In the above expression of $U\left( C\right) $, $%
\left\vert a_{0}\right\rangle _{2}\left\langle a_{0}\right\vert $ is the
projection operator of the second qubit to the state $\left\vert
a_{0}\right\rangle _{2}$. It is easy to see that the holonomy $U\left(
C\right) $ is the production of a control phase gate operation and a
single-qubit rotation about the $z$ axis operated on the second qubit. When $%
\phi \left( C\right) \neq 0$, $U\left( C\right) $ is an nontrivial two-qubit
operation. If we choose the path $C$ which satisfing $\eta \left( C\right)
=0 $, we can obtain an explicit control phase gate via the holonomy $U\left(
C\right) =$ $e^{i\phi \left( C\right) \left\vert a_{0}a_{0}\right\rangle
\left\langle a_{0}a_{0}\right\vert }$.

Moreover, we may also realize the controlled-NOT by setting $\theta ^{\left(
1\right) }=\phi _{0}^{\left( j\right) }=\phi _{1}^{\left( j\right) }=0$ and
choose $\theta ^{\left( 2\right) }$, $\xi ^{\left( 1\right) }$ and $\xi
^{\left( 2\right) }$ as the control parameters. In this case, the dark
states can also be obtained with some straightforward derivations, but have
quite complicated forms (not presented here). Our main result is that, when $%
-\int_{C}\sin \left[ \theta ^{\left( 2\right) }\right] d\xi ^{\left(
2\right) }-\theta \left( C\right) =\pi /2$, the holonomy associated with a
closed path can be expressed as $U\left( C\right) =e^{-i\frac{\pi }{4}%
}R_{y}^{\left( 2\right) }\left[ \theta \left( C\right) \right] R_{z}^{\left(
1\right) }\left( \frac{\pi }{4}\right) R_{z}^{\left( 2\right) }\left( \frac{%
\pi }{4}\right) U_{CN}R_{z}^{\left( 2\right) }\left( -\frac{\pi }{4}\right) $%
, where%
\begin{equation}
U_{CN}=\left\vert a_{0}\right\rangle _{1}\left\langle a_{0}\right\vert
\otimes I^{\left( 2\right) }+\left\vert a_{1}\right\rangle _{1}\left\langle
a_{1}\right\vert \otimes \sigma _{x}^{\left( 2\right) }  \nonumber
\end{equation}%
is just the controlled-NOT operation. Here, $I^{\left( 2\right) }$ is the
identity operator of the second qubit and the angle $\theta \left( C\right) $
are defined as $\theta \left( C\right) =-\int_{C}\beta ^{-1}\Lambda
_{01}^{-1}\Lambda _{00}^{-3}\Lambda _{2}d\xi ^{\left( 2\right)
}-\int_{C}\beta ^{-1}\Lambda _{01}^{-1}\Lambda _{00}^{-3}\Lambda _{1}d\xi
^{\left( 1\right) }$, where $\Lambda _{00}$ is defined as before and the
other coefficients are defined as $\Lambda _{01}=\sqrt{2-\sin ^{2}\left[ \xi
^{\left( 1\right) }\right] \cos ^{2}\left[ \xi ^{\left( 2\right) }\right]
\sin ^{2}\left[ \theta ^{\left( 2\right) }\right] }$, $\alpha =-\left(
1/2\right) \Lambda _{00}^{-1}\Lambda _{01}^{-1}\sin ^{2}\left[ \xi ^{\left(
1\right) }\right] \sin \left[ 2\xi ^{\left( 2\right) }\right] \sin \left[
\theta ^{\left( 2\right) }\right] $ and $\beta =\sqrt{1+\alpha ^{2}}.$
Therefore, up to a global phase factor, $U\left( C\right) $ is just the
product of control not operation and some single-qubit rotations.
%Then we can realize a control not gate by the holonomic associated
%from the path $C$ which satisfies $\zeta \left( C\right) =\pi /2$,
%together with some single-qubit rotations.

In most cases, the Hamiltonian (\ref{j}) has a four-dimension engenspace
with zero eigenvalue in the invariant subspace $I$. The basis of the
eigenspace are just the dark states $\left\vert D_{ij}\right\rangle $.
Nevertheless, it is pointed out that, in some very special cases, there may
be accidental degeneracy in a sub-manifold (we call it the AD sub-manifold)
of the parameter space $M$. In the AD sub-manifold, in addition to the dark
states $\left\vert D_{ij}\right\rangle $, the Hamiltonian has another two
eigenstates with zero eigenvalue and thus the dimension of the Hamiltonian's
eigenspace with zero eigenvalue is six rather than four. It is apparently
that if the path of the adiabatic evolution of the Rabi frequencies cross
the AD manifold, there might be a transition from the four dark states $%
\left\vert D_{ij}\right\rangle $ to the two external states. To avoid this
kind of unwanted transition, we should control the evolution path of the
Rabi frequencies in the parameter space $M$ be far away enough from the AD
sub-manifold. On the other hand, since the evolution of the Rabi frequencies
in $M$ is assumed to begin and end at the same point $O$ where all the Rabi
frequencies are set to zero, the accidental degeneracy at $O$ should be
avoided. This can be implemented by adjusting the coupling strength $%
g^{\left( l\right) }$ via controlling the position of the SQUIDs in the
cavity, or the detuning $\Delta ^{\left( l\right) }$ of each SQUID via
changing the bias fields. For instance, if the conditions $\Delta ^{\left(
1\right) }=\Delta ^{\left( 2\right) }\neq 0,g^{\left( 1\right) }=g^{\left(
2\right) }$ or $\Delta ^{\left( 1\right) }=\Delta ^{\left( 2\right)
}=0,g^{\left( 1\right) }\neq g^{\left( 2\right) }$ are satisfied, the
accidental degeneracy at point $O$ can be avoided.

Since the dark states have a non-zero projection to the single or two photon
states, the photon dissipation caused by the imperfection would tend to
destroy the dark states. To evaluate the influence of the photon
dissipation, the Schoredinger equation controlled by the effective
Hamiltonian
\begin{equation}
H_{eff}=H\left( t\right) -i\left( \kappa /2\right) a^{\dagger }a  \label{aa}
\end{equation}%
needs to be solved. Here, $H\left( t\right) $ is the Hamiltonian defined in
Eq. (\ref{j}) (or Eq. (\ref{b})), and $\kappa $ the photon decay rate. The
dissipation term $-i\left( \kappa /2\right) a^{\dagger }a$ can be considered
as a perturbation. As a result of the first order perturbation theory, the
fidelity of the quantum gate operation, i.e., the probability of the ideal
finial state, may be expressed as
\begin{equation}
F\approx 1-\kappa \int_{0}^{T}\left\langle n\right\rangle _{ph}dt  \label{bb}
\end{equation}%
where $T$ is the operation time and $\left\langle n\right\rangle
_{ph}=\left\langle \Psi \left( t\right) \right\vert a^{\dagger }a\left\vert
\Psi \left( t\right) \right\rangle $ is the instantaneous expectation value
of photon number. Therefore, the condition under which the influence of
photon dissipation can be neglected is simply $\kappa
\int_{0}^{T}\left\langle n\right\rangle _{ph}dt<<1$. On the other hand,
since our scheme is based on the adiabatic evolution of the quantum states,
the adiabatic condition should be satisfied, which can be expressed as $%
\tilde{\Omega}T>>1$, where $\tilde{\Omega}$ is the energy gap between the
dark state and other eigenstates of the Hamiltonian \cite{gate1}. Here, $%
\tilde{\Omega}$ has the same order of amplitude as the SQUID-cavity coupling
constant $g$. In practical quantum gate operations, we always have $T\sim
10^{3}g^{-1}$. Then the condition $\kappa \int_{0}^{T}\left\langle
n\right\rangle _{ph}dt<<1$ can be satisfied when $g/\kappa \gtrsim 10^{4}$.
The coupling constant of the SQUID and the cavity available at present is $%
g\sim 1.8\times 10^{8}s^{-1}$ \cite{han2}. The high quality factor of the
cavity $Q=10^{6}\sim 10^{8}$ might be achieved experimentally \cite{cavity}.
This will lead to $\kappa \int_{0}^{T}\left\langle n\right\rangle
_{ph}dt\lesssim 10^{-2}$ and thus the fidelity $F\simeq 1$.
\begin{figure}[h]
\begin{center}
\includegraphics[width=4.5cm,height=3.5cm]{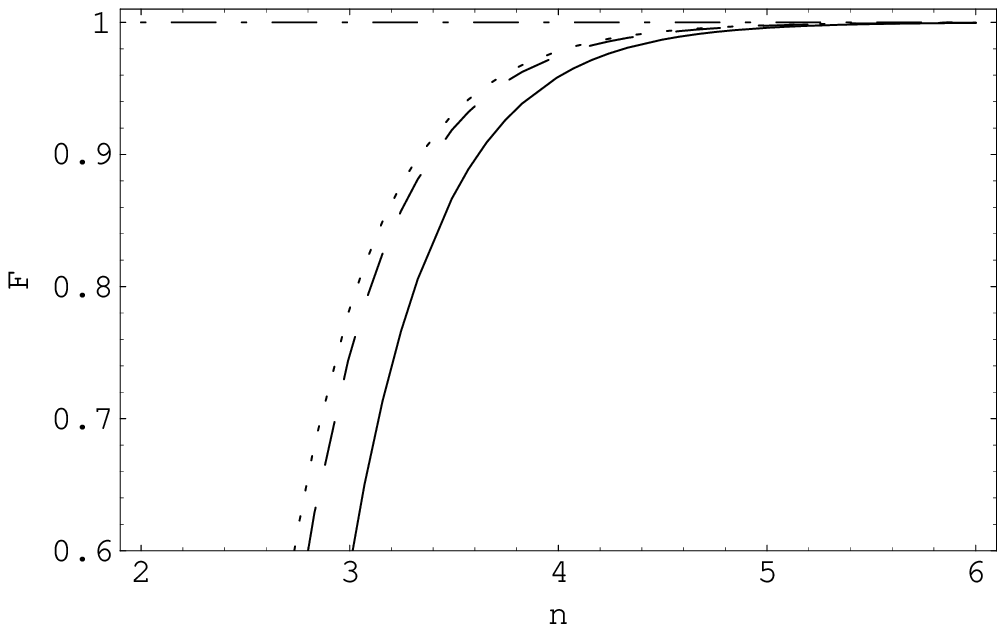} \vspace{0.3cm}
\end{center}
\caption{ The fidelities of a control phase gate for the initial states $%
\left\vert a_{0}\right\rangle _{1}\left\vert a_{0}\right\rangle _{2}$ (solid
line), $\left\vert a_{0}\right\rangle _{1}\left\vert a_{1}\right\rangle _{2}$
(dashed line), $\left\vert a_{1}\right\rangle _{1}\left\vert
a_{0}\right\rangle _{2}$ (dotted line), and $\left\vert a_{1}\right\rangle
_{1}\left\vert a_{1}\right\rangle _{2}$ (dashed-dotted line), where the
quantity $n$ is defined by the relation $g/\protect\kappa =10^{n}$. }
\end{figure}

Finally, let us look into in some detail a typical control phase gate
operation discussed before. In this operation, we assume $g^{\left( 1\right)
}=2g^{\left( 2\right) }=g=1.8\times 10^{8}s^{-1}$ and $\Delta ^{\left(
1\right) }=\Delta ^{\left( 2\right) }=0$. The amplitudes of the Rabi
frequencies $\Omega _{0}^{\left( 1\right) }$ and $\Omega _{0}^{\left(
2\right) }$ are varied following the Gaussian functions of time: $\Omega
_{0}^{\left( 1\right) }=2.5ge^{-\left( \frac{t-3\tau }{\tau }\right)
^{2}},\Omega _{0}^{\left( 2\right) }=ge^{-\left( \frac{t-3\tau }{\tau }%
\right) ^{2}}e^{i\phi _{0}^{\left( 2\right) }\left( t\right) }$, where $\tau
=144g^{-1}$. The phase $\phi _{0}^{\left( 2\right) }\left( t\right) $ is set
to be a hyperbolic tangent function of time: $\phi _{0}^{\left( 2\right)
}\left( t\right) =\pi \times \left[ 1+\text{tanh}\left( \frac{t}{0.75\tau }%
\right) \right] $. As in the above discussions, the control gate operation
can be written as an unitary transformation $U=e^{i\eta \left\vert
a_{0}\right\rangle _{2}\left\langle a_{0}\right\vert }e^{i\phi \left\vert
a_{0}a_{0}\right\rangle \left\langle a_{0}a_{0}\right\vert }$, where $\phi
\approx \pi /6$ and $\eta \approx 4$. The operation time of this gate is
about $8\times 10^{2}g^{-1}$. We estimate the fidelity of this operation
using Eq. (\ref{bb}). In Fig. 2, the fidelity of the quantum gate operation
with four possible initial states are plotted as a function of the ratio $%
g/\kappa $. It is seen that when $g/\kappa \sim 10^{3}-10^{4}$, the fidelity
is larger than $90\%$. In particular, whenever $g/\kappa \sim 10^{5}$, the
fidelity is improved to reach $99\%$.

We thank P. Zanardi and Y. Li for useful discussions. We also
thank Prof. Siyuan-Han for his grate suggestion on the utilization
of rf-SQUID whose effective potential is triple well. The work was
supported by the RGC grant of Hong Kong (HKU7114/02P), the CRCG
grant of HKU, the NSFC, the knowledge Innovation Program (KIP) of
the Chinese Academy of Sciences, and the National Fundamental
Research Program of China (001CB309310).

\end{document}